\documentclass[aps,prl,twocolumn,10pt,showpacs]{revtex4}
\usepackage{amssymb,amsthm,amsmath,amsfonts}
\usepackage{color,hyperref,graphicx,ulem,pdfpages}

\begin{document}
 
\title{Uncovering Quantum Correlations with Time-Multiplexed Click Detection}

\author{J. Sperling}\affiliation{Arbeitsgruppe Theoretische Quantenoptik, Institut f\"ur Physik, Universit\"at Rostock, D-18051 Rostock, Germany}
\author{M. Bohmann}\affiliation{Arbeitsgruppe Theoretische Quantenoptik, Institut f\"ur Physik, Universit\"at Rostock, D-18051 Rostock, Germany}
\author{W. Vogel}\affiliation{Arbeitsgruppe Theoretische Quantenoptik, Institut f\"ur Physik, Universit\"at Rostock, D-18051 Rostock, Germany}

\author{G. Harder}\affiliation{Integrated Quantum Optics Group, Applied Physics, University of Paderborn, 33098 Paderborn, Germany}
\author{B. Brecht}\affiliation{Integrated Quantum Optics Group, Applied Physics, University of Paderborn, 33098 Paderborn, Germany}
\author{V. Ansari}\affiliation{Integrated Quantum Optics Group, Applied Physics, University of Paderborn, 33098 Paderborn, Germany}
\author{C. Silberhorn}\affiliation{Integrated Quantum Optics Group, Applied Physics, University of Paderborn, 33098 Paderborn, Germany}
\date{\today}

\begin{abstract}
	We report on the implementation of a time-multiplexed click detection scheme to probe quantum correlations between different spatial optical modes.
	We demonstrate that such measurement setups can uncover nonclassical correlations in multimode light fields even if the single mode reductions are purely classical.
	The nonclassical character of correlated photon pairs, generated by a parametric down-conversion, is immediately measurable employing the theory of click counting instead of low-intensity approximations with photoelectric detection models.
	The analysis is based on second- and higher-order moments, which are directly retrieved from the measured click statistics, for relatively high mean photon numbers.
	No data postprocessing is required to demonstrate the effects of interest with high significance, despite low efficiencies and experimental imperfections.
	Our approach shows that such novel detection schemes are a reliable and robust way to characterize quantum-correlated light fields for practical applications in quantum communications.
\end{abstract}
\maketitle


\paragraph{Introduction.--}
	Certifying quantum features of light is one key requirement for optical implementations of quantum-information technology~\cite{KMNRDM07,DZTDSW11}.
	This requires, on the one hand, reliable sources for correlated quantum light and, on the other hand, appropriate detection schemes~\cite{BC10,EFMP11}.
	Since correlations between different degrees of freedom can have different origins in quantum optics, they might be covered by classical statistical optics, or they are genuine quantum properties having no such classical counterpart.
	Using the classical theory of coherence, one way to discern quantum from classical effects has been introduced independently by Glauber~\cite{G63} and Sudarshan~\cite{S63}.

\begin{figure}
		\centering
		\includegraphics[width=7cm]{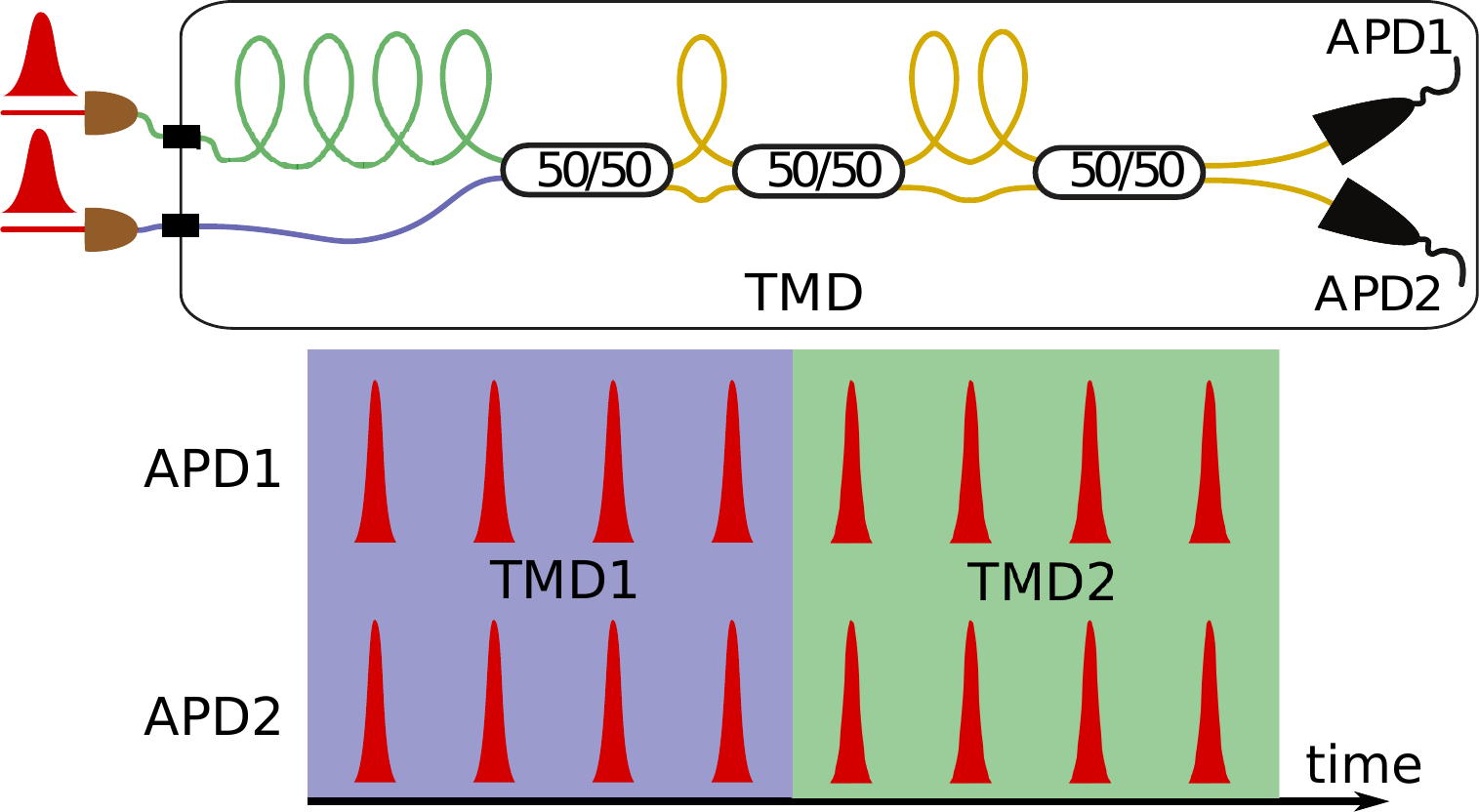}
		\caption{
			(Color online) The realization of two TMDs in a single device is depicted.
			One part (upper, green input) of the correlated signal field is delayed in an optical fiber before entering the device.
			Within the TMD, any signal is split by a 50/50 beam splitter followed by another delay line in one output.
			This multiplexing is done three times yielding $2^3$ separated time bins for each (upper green and lower blue) input field which are measured with two APDs.
		}\label{Fig:TMDsetup}
	\end{figure}

	A quantum-state characterization is often based on the photon number distribution, see, e.g., Refs.~\cite{MW95,A13}.
	The corresponding photoelectric detection theory yields Poissonian statistics for coherent light.
	However, detectors that directly measure photon numbers are typically not available or require advanced data postprocessing~\cite{SSG09,HPHP05}.
	Today, quantum states with low mean photon number are often detected with avalanche photodiodes (APDs) in the Geiger mode, which basically produce a ``click'' if any number of photons is absorbed, and remain silent otherwise.
	A uniform splitting of a radiation field with many photons into portions of lower intensities, each being measured with an APD, can extent the knowledge about the signal, for example, to discriminate single- and two-photon events.
	Various kinds of such photon-number-resolving detectors have been implemented to demonstrate nonclassical features of radiation fields, e.g., in Refs.~\cite{ABA10,PDFEPW10,WDSBY04,DBJVDBW14,MMDL12,DYSTS11,DBIMHCE13}, or for characterizing the non-Poissonian behavior of the click statistics~\cite{BDFL08,FJPF03}.
	One particular realization of such a scheme that requires only a small number of optical elements is so-called time-bin multiplexing detectors (TMDs)~\cite{FJPF03,ASSBW03,RHHPH03}, cf. Fig.~\ref{Fig:TMDsetup}.
	Based on these TMDs, one can reconstruct nonclassical features of quantum light fields~\cite{LCGS10,HSRHMSS14} or higher-order correlation functions~\cite{ALCS10}.

	The proper theoretical detection model for such devices is a quantum version of the binomial statistics~\cite{SVA12}.
	It was also shown that approximating these statistics with a Poissonian distribution and applying quantumness probes of the photon statistics can yield fake signatures of nonclassicality for classical fields -- even if the number of registered signal photons is one order of magnitude below the number of time bins.
	To eliminate these errors, nonclassicality criteria in terms of moments of the click-counting statistics have been proposed to directly uncover quantum light with click counters without data postprocessing~\cite{SVA13}.
	For example, the notion of sub-binomial light has been introduced~\cite{SVA12a} and experimentally demonstrated~\cite{BDJDBW13,HSPGHNVS15}.

	In the present work, we report on the characterization of a bipartite quantum-correlated light source using click detectors only.
	We demonstrate that click detection is capable of directly verifying nonclassicality even for imperfect experimental settings.
	For our parametric down-conversion (PDC) based light source, this approach correctly shows classical single-mode correlations and, at the same time, it uncovers quantum correlations between the modes.
	We show that our simple treatment to infer these quantum features works for a broad range of pump powers.
	In particular, it works increasingly well for increasing pump powers, where the often used weak signal approximations with the photoelectric detection theory completely break down, as we will show later.

\paragraph{Nonclassical moments of the click statistics.--}
	The probability to measure $k_A$ clicks within the $N=8$ time bins assigned to the signal $A$ together with $k_B$ clicks from the signal $B$, cf. Fig.~\ref{Fig:TMDsetup}, is described through the joint click-counting statistics~\cite{SVA12,SVA13}:
	\begin{align}\label{Eq:JointClickCounting}
	\begin{aligned}
		c_{k_A,k_B}=&\langle{:} \binom{N}{k_A}\hat m_A^{N-k_A}(\hat 1_A-\hat m_A)^{k_A}\\
					&\times\binom{N}{k_B}\hat m_B^{N-k_B}(\hat 1_B-\hat m_B)^{k_B}{:}\rangle,
	\end{aligned}
	\end{align}
	with ${:}\,\cdot\,{:}$ denoting the normally ordering prescription and $\hat m_i=e^{-\Gamma(\hat n_i/N)}$ for the modes $i=A,B$.
	In general, $\Gamma$ can be an unknown detector response, being a function of the photon number operators $\hat n_i$.
	For example, a linear form of the response function is $\Gamma(\hat n_i/N)=\eta\hat n_i/N+\nu$, with $\eta$ and $\nu$ being the quantum efficiency and the dark count rate, respectively.

	In Ref.~\cite{SVA13} it has been demonstrated that the matrix of click moments $M^{(K_A,K_B)}$ is non-negative for any classical light field,
	\begin{align}\label{Eq:ClMoM}
		0\leq M^{(K_A,K_B)}{=}(\langle{:}\hat m_A^{s_A{+}t_A}\hat m_B^{s_B{+}t_B}{:}\rangle )_{(s_A,s_B),(t_A,t_B)},
	\end{align}
	with $s_i,t_i=0,\ldots, K_i/2\leq N/2$ for even $K_i$ and $N$.
	The superscript $(K_A,K_B)$ defines the highest moment of each subsystem within the matrix $M$, see also Supplement.
	For instance, the single-mode and bipartite, second-order matrices of click moments are
	\begin{align}\label{Eq:SecondOrderMoM}
		&M^{(2,0)}{=}\!\begin{pmatrix} 1 & \langle{:}\hat m_A{:}\rangle \\ \langle{:}\hat m_A{:}\rangle & \langle{:}\hat m_A^2{:}\rangle \end{pmatrix}\!\!,\text{ }
		M^{(0,2)}{=}\!\begin{pmatrix} 1 & \langle{:}\hat m_B{:}\rangle \\ \langle{:}\hat m_B{:}\rangle & \langle{:}\hat m_B^2{:}\rangle \end{pmatrix}\!\!,\nonumber\\
		&\text{and }M^{(2,2)}{=}\!\begin{pmatrix}
			1 & \langle{:}\hat m_A{:}\rangle & \langle{:}\hat m_B{:}\rangle \\
			\langle{:}\hat m_A{:}\rangle & \langle{:}\hat m_A^2{:}\rangle & \langle{:}\hat m_A\hat m_B{:}\rangle \\
			\langle{:}\hat m_B{:}\rangle & \langle{:}\hat m_A\hat m_B{:}\rangle  & \langle{:}\hat m_B^2{:}\rangle
		\end{pmatrix},
	\end{align}
	respectively.
	The needed moments can be directly retrieved from the measured click-counting statistics~\cite{SVA13},
	\begin{align}\label{Eq:SamplingFormula}
		\langle{:}\hat m_A^{l_A}\hat m_B^{l_B}{:}\rangle=\sum_{k_A=0}^{N-l_A}\sum_{k_B=0}^{N-l_B}
		\frac{\binom{N-k_A}{l_A}\binom{N-k_B}{l_B}}{\binom{N}{l_A}\binom{N}{l_B}}c_{k_A,k_B}.
	\end{align}

	One way to probe the character of nonclassical correlations, i.e., violating inequality~\eqref{Eq:ClMoM}, can be done as follows.
	We have nonclassical $K$th-order click correlations if
	\begin{align}\label{Eq:KthOrderCorrelations}
		M^{(K,0)}\geq 0,\text{ }M^{(0,K)}\geq 0,\text{ and }M^{(K,K)}\ngeq 0.
	\end{align}
	This means that both $K$th-order single-mode marginals are classical and the bimodal, $K$th-order correlation matrix is nonclassical, i.e., it has at least one negative eigenvalue.
	In order to genuinely certify such nonclassical correlations, it is sufficient to consider the minimal eigenvalues of the click-moment matrices.
	Say $\vec f_A$, $\vec f_B$, and $\vec f_{AB}$ are the normalized eigenvectors to the minimal eigenvalues $e_A$, $e_B$, and $e_{AB}$ of $M^{(K,0)}$, $M^{(0,K)}$, and $M^{(K,K)}$, respectively.
	Now, definition~\eqref{Eq:KthOrderCorrelations} is rewritten as
	\begin{align}\label{Eq:MinEval}
	\begin{aligned}
		e_A=\vec f_A^{\,\dagger} M^{(K,0)}\vec f_A\geq 0,\,
		e_B=\vec f_B^{\,\dagger} M^{(0,K)}\vec f_B\geq 0,\\
		\text{and}\, e_{AB}=\vec f_{AB}^{\,\dagger} M^{(K,K)}\vec f_{AB}<0.
	\end{aligned}
	\end{align}
	This method will serve as our approach to determine $K$th-order quantum correlations between the subsystems $A$ and $B$, see Supplement for further details.

\paragraph{Implementation and model.--}
	The states under study are produced in a type II PDC in a periodically poled, $8\,{\rm mm}$ long KTP waveguide.
	The PDC process is pumped with $1\,\textrm{ps}$ long pulses at a repetition rate of $70\,\textrm{kHz}$ and a wavelength of $768\,\textrm{nm}$ coming from a Ti:Sapphire laser.
	The states are generated in two orthogonally polarized signal and idler modes at $1536\,\textrm{nm}$.
	In the regime of less than one photon per pulse, the source has been characterized in Ref.~\cite{harder13} with emphasis on the modal properties, and genuine single-mode operation of the PDC has been shown.
	Behind the waveguide, broadband spectral filters are used to suppress the pump and unwanted background outside of the PDC spectral region.
	Signal and idler modes are then split at a polarizing beam splitter and coupled into a two-mode TMD consisting of a fiber network with 50/50 beam splitters and a pair of InGaAs APDs as depicted in Fig.~\ref{Fig:TMDsetup}.
	For each pump setting, we record the full time tagging data for $10\,{\rm min}$ and translate them into the click statistics of all possible $8\times 8$ click events.
	Experimental imperfections are unbalanced beam splitters within our TMD, i.e., splitting ratios that slightly deviate from $50/50$, or after-pulsing effects~\cite{kang03}.
	We minimized both effects through a careful setup preparation.

\begin{figure}
		\centering
 		\includegraphics[width=7cm]{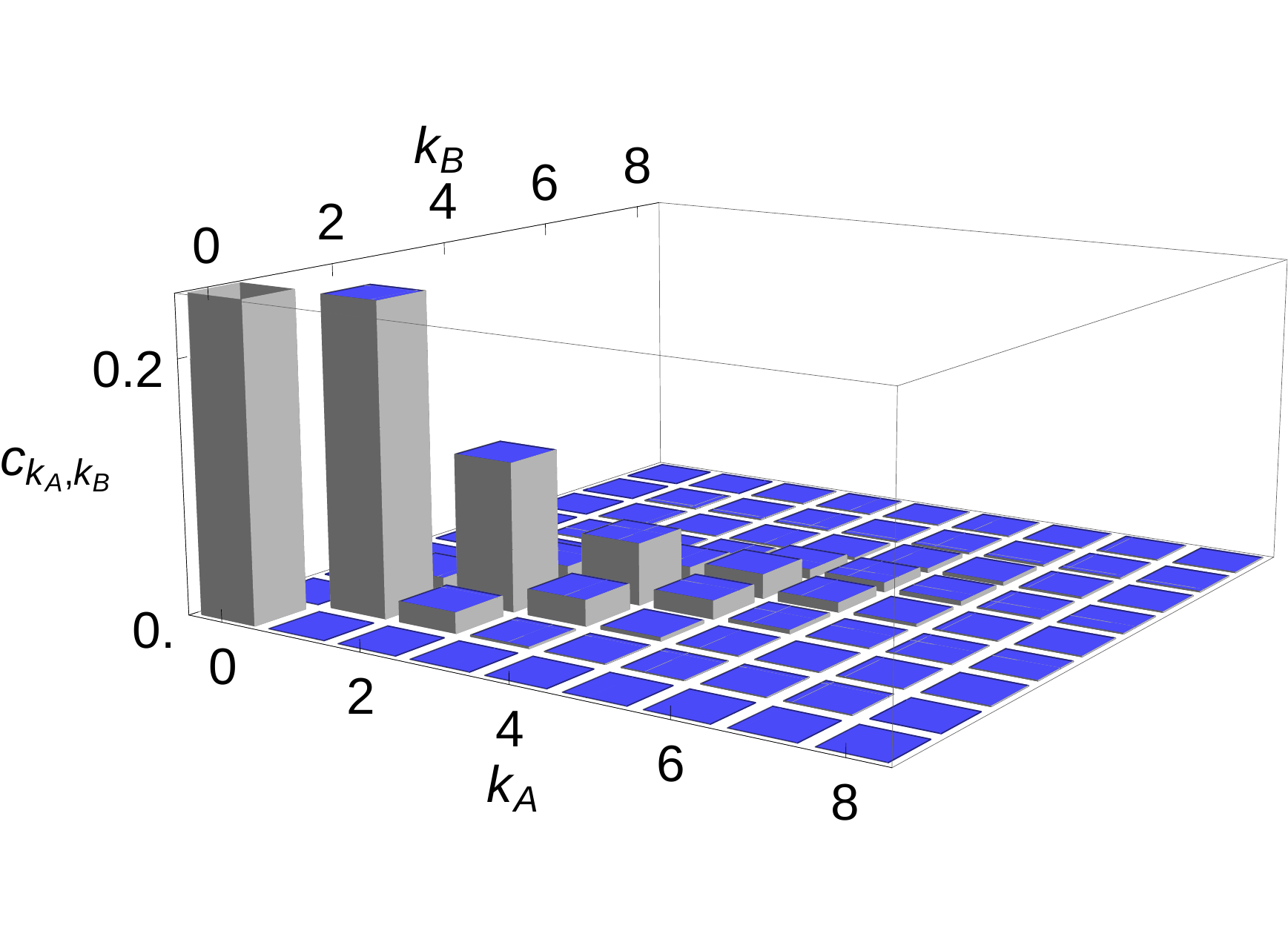}
		\caption{(Color online)
			The joint click-counting statistics~\eqref{Eq:JointClickCounting} for the state~\eqref{Eq:TMSV}, $\xi{=}1$, and a perfect linear response, $\Gamma(\hat n/N){=}\hat n/N$.
			The probability for no clicks is cut, $c_{0,0}{=}0.42$.
			Even though we have perfect photon correlations, the click-counting distribution includes off-diagonal terms.
		}\label{Fig:Example}
	\end{figure}

	The PDC may be formulated in terms of the effective Hamiltonian
	$\hat H=i\kappa\gamma\hat a^\dagger\hat b^\dagger+{\rm H.c.}$, where $\kappa$ is a coupling constant, $\gamma$ is the coherent amplitude of the pump beam, and $\hat a^\dagger(\hat b^\dagger)$ is the creation operator of the mode $A(B)$.
	An idealized, perfect unitary evolution for this process yields the two-mode squeezed-vacuum state,
	\begin{align}\label{Eq:TMSV}
		|\xi\rangle=e^{\xi(\hat a^\dagger\hat b^\dagger-\hat a\hat b)}|{\rm vac}\rangle
		=\sum_{n=0}^\infty \frac{(\tanh\xi)^n}{\cosh \xi} |n\rangle_A|n\rangle_B,
	\end{align}
	where $\xi\geq0$ is proportional to the square root of the pump power $P$, since $\hat H\propto\gamma$ and $P\propto \gamma^2$.
	The noise suppression of the squeezed quadrature $\hat X$ is $\langle [\Delta \hat X]^2\rangle=e^{-2\xi}\langle [\Delta \hat X]^2\rangle_{\rm vac}$~\cite{A13} and Supplement.
	Because of the pairwise generation of photons, the state~\eqref{Eq:TMSV} has perfect photon-number correlations.
	This means that whenever a certain number of photons is present in one mode, the same number of photons occurs in the other mode.
	However, the click-counting statistics include off-diagonal elements $c_{k_A,k_B}\neq0$ for $k_A\neq k_B$, even for a perfect detection without dark counts and unit efficiency, cf. Fig.~\ref{Fig:Example}.
	This difference between click-counting and photoelectric detection is due to a finite probability that more than one photon can end in the same time bin.
	It is worth mentioning that the single-mode reduced states ${\rm tr}_A|\xi\rangle\langle\xi|$ and ${\rm tr}_B|\xi\rangle\langle\xi|$ are classical thermal states, and that the total number of photons is $\bar n=\langle \xi|\hat n_A+\hat n_B|\xi\rangle=2\sinh^2\xi$.

\begin{figure}
		\centering
		\includegraphics[width=8cm]{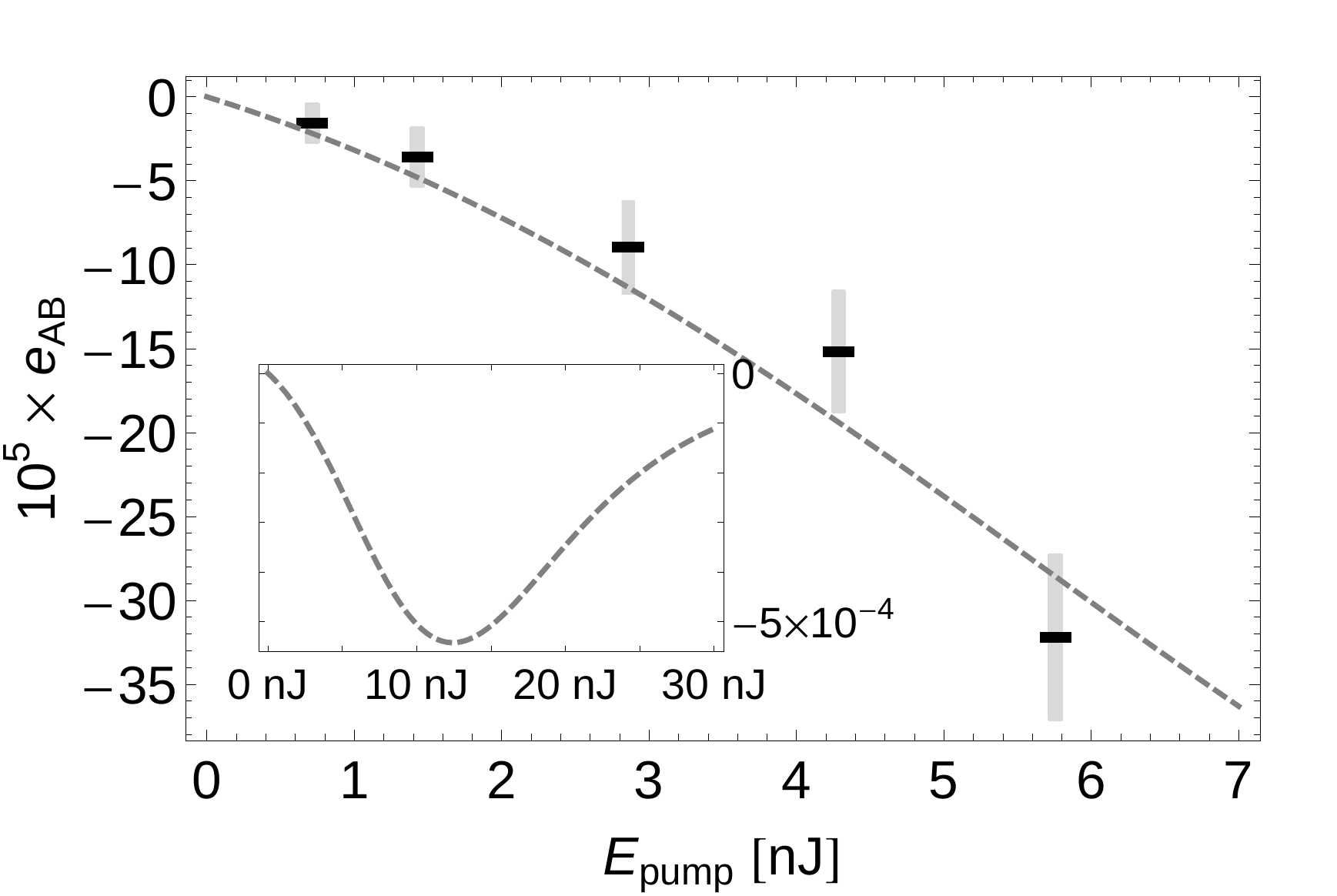}\\[0.5cm]
		\includegraphics[width=4cm]{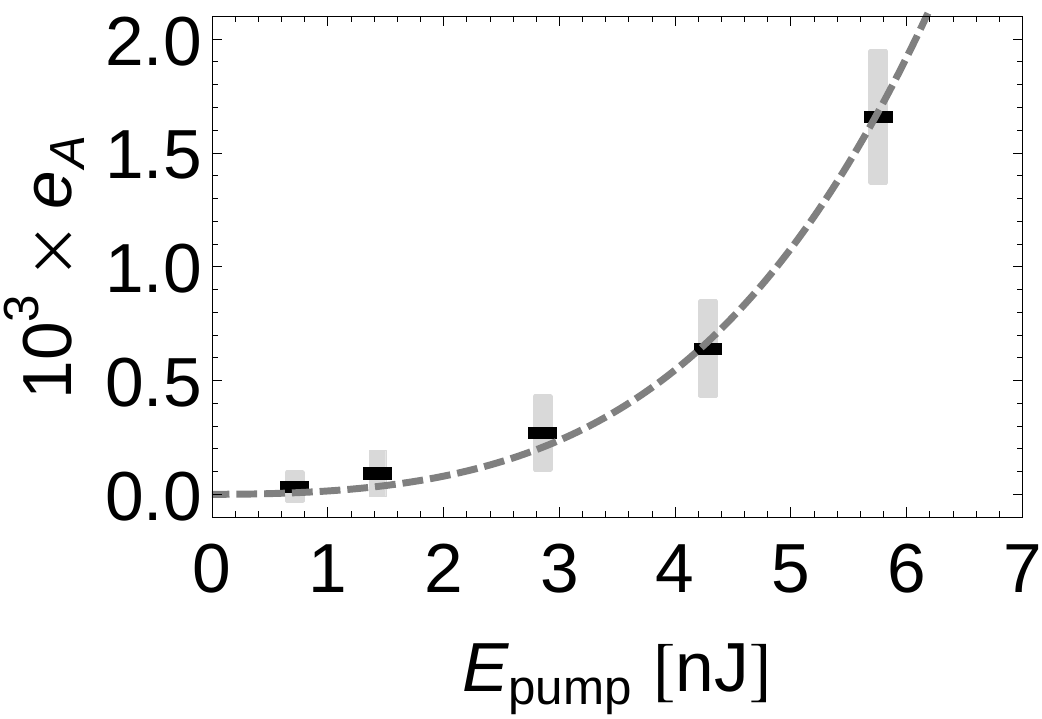}\hspace*{0.5cm}
		\includegraphics[width=4cm]{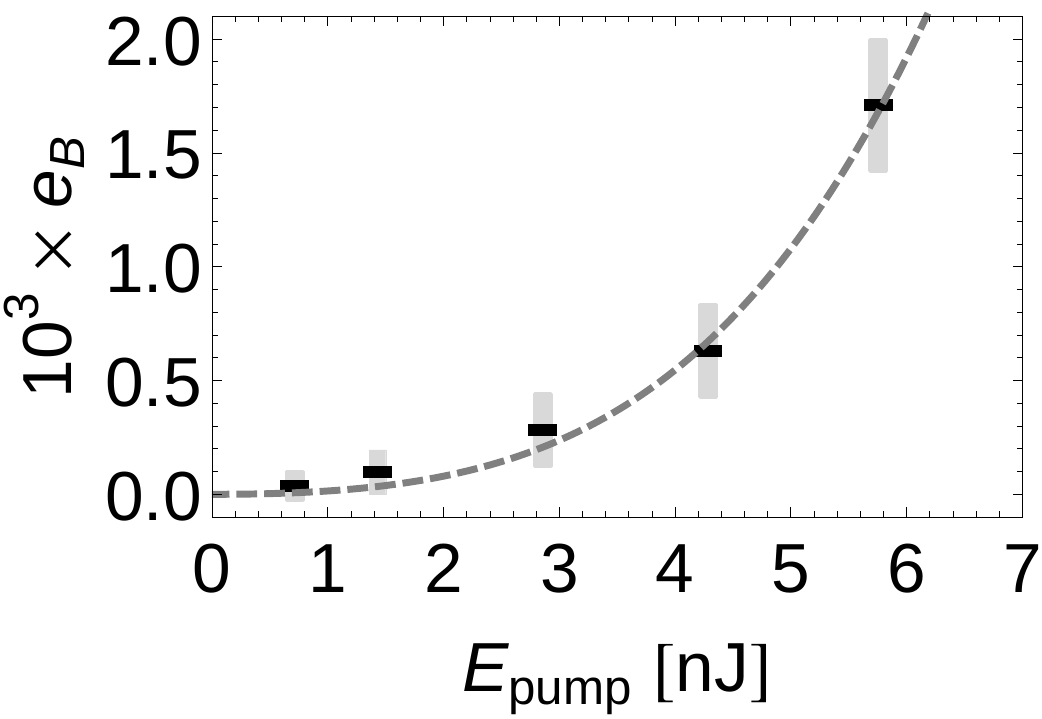}
		\caption{
			The upper plot shows the minimal eigenvalue $e_{AB}$ (black bars) of the experimentally obtained matrix of moments $M^{(2,2)}$ depending on the pump energy $E_{\rm pump}$.
			Error bars are given as gray areas.
			The dashed curve is the theoretical prediction.
			The inset depicts the continuation of the theoretical curve for higher energies including saturation effects.
			The lower plots show the minimal eigenvalues $e_A$ and $e_B$ of $M^{(2,0)}$ (left) and $M^{(0,2)}$ (right), respectively, a ten standard deviations error bar, and the theoretical prediction.
			Since the bimodal correlations are negative and the single-mode reductions are non-negative, we successfully determined nonclassical correlations between the modes $A$ and $B$.
		}\label{Fig:2ndOrder}
	\end{figure}

\paragraph{Second-order correlations.--}
	First, let us focus on second-order click correlations, cf. Eq.~\eqref{Eq:SecondOrderMoM}, which include the information about the mean values, the variances, and the covariance of the joint click-counting statistics~\cite{SVA13}.
	In Fig.~\ref{Fig:2ndOrder}, we plot the measurement results.
	Using the approach in Eq.~\eqref{Eq:MinEval}, the minimal eigenvalues $e_{AB}$ (top), $e_A$ (bottom, left), and $e_B$ (bottom, right) are shown in their dependence on the energy per pulse, $E_{\rm pump}=P/70\,{\rm kHz}$.
	The energy can be experimentally controlled.
	The single-mode matrices are non-negative, $e_A\geq0$ and $e_B\geq0$, whereas the cross correlations are nonclassical, $e_{AB}<0$.
	Thus, we have verified the quantum nature of the second-order click correlations between the spatial modes $A$ and $B$.

	In order to compare our measured results, a simple theoretical model is used.
	We assume that the pure state~\eqref{Eq:TMSV} is generated, and the detectors are described via a plain linear response function: $\Gamma(\hat n/N)=\eta\hat n/N+\nu$.
	As we discussed above, the parameter $\xi$, characterizing the state $|\xi\rangle$, depends on the pump power $P$: $\xi=\xi_0\sqrt P$.
	The proportionality constant $\xi_0$, the quantum efficiency $\eta$, and the dark count rate have been fitted using the standard method of least squares.
	They are $\eta=9.6\%$, $\nu=0.51$, and $\xi_0=0.087\,({\rm \mu W})^{-1/2}$.
	With these values, we estimate the total mean photon numbers to be $0.9\ldots15$ photons for the pump powers $50\ldots403\,{\rm \mu W}$ or energies $0.7\ldots5.8\,{\rm nJ}$.
	Already this simplified model yields a good agreement with the measured results, which highlights the excellent performance of the engineered PDC source, cf. the dashed lines in Fig.~\ref{Fig:2ndOrder}.

	The inset in the upper plot shows the extrapolation of the quantum correlations, $e_{AB}<0$, for higher pump energies.
	At some point, the mean photon number is so high, that these correlations saturate and eventually vanish.
	A high squeezing level and a classical laser light with a large coherent amplitude result in the same signal -- all time bins are occupied with a large number of photons.
	Therefore, at high photon numbers, the signals of nonclassical and classical states cannot be discriminated.
	Thus, the recorded quantum correlations of the former state must decrease due to the saturation of click-counting devices, which is automatically included in the click-counting theory.
	Imperfections are also studied in the Supplement.

	Let us emphasize again that the states have been generated for pump powers ranging over almost one order of magnitude.
	Verifying nonclassical photon-photon correlations in this comparably large domain is typically considered a challenging task, but can easily be accomplished with our TMD click counters.
	In addition, since the method of nonclassical click moments is independent of the state, the verified quantum correlations can be certified even if the pump power was completely unknown.

\paragraph{Higher-order correlations.--}
	Let us study higher-order quantum correlations.
	They become particularly meaningful when second-order criteria fail to properly characterize the state~\cite{AT92}.
	For example, the third- and fourth-order moment relate to the so-called skewness and the flatness (or kurtosis), respectively.
	The highest possible order of moments one can infer from $N=8$ time bins per mode is $K=8$ in Eqs.~\eqref{Eq:ClMoM} and~\eqref{Eq:MinEval}, which yields a full characterization of the click-counting statistics.

	The bound for a classical signal, $M^{(K,K)}\geq0$, is given by the eigenvalue $e_{\rm cl}=0$~\cite{SVA13}.
	Thus, the signed distance, in units of standard deviations, of the experimentally obtained minimal eigenvalue $e=\overline e\pm\Delta e$ to this classical bound leads to a signed significance,
	\begin{align}
		\Sigma=\frac{\overline e-e_{\rm cl.}}{\Delta e}=\frac{\overline e}{\Delta e},
	\end{align}
	representing a signed relative error.
	A negative significance $\Sigma<0$ verifies the nonclassicality with a significance of $|\Sigma|$.
	Typically, $\Sigma\lesssim-3$ is a significant verification of the negativity, whereas $\Sigma\approx 0$ cannot be distinguished from the classical bound $0$.

\begin{figure}
		\centering
		\includegraphics[width=7cm]{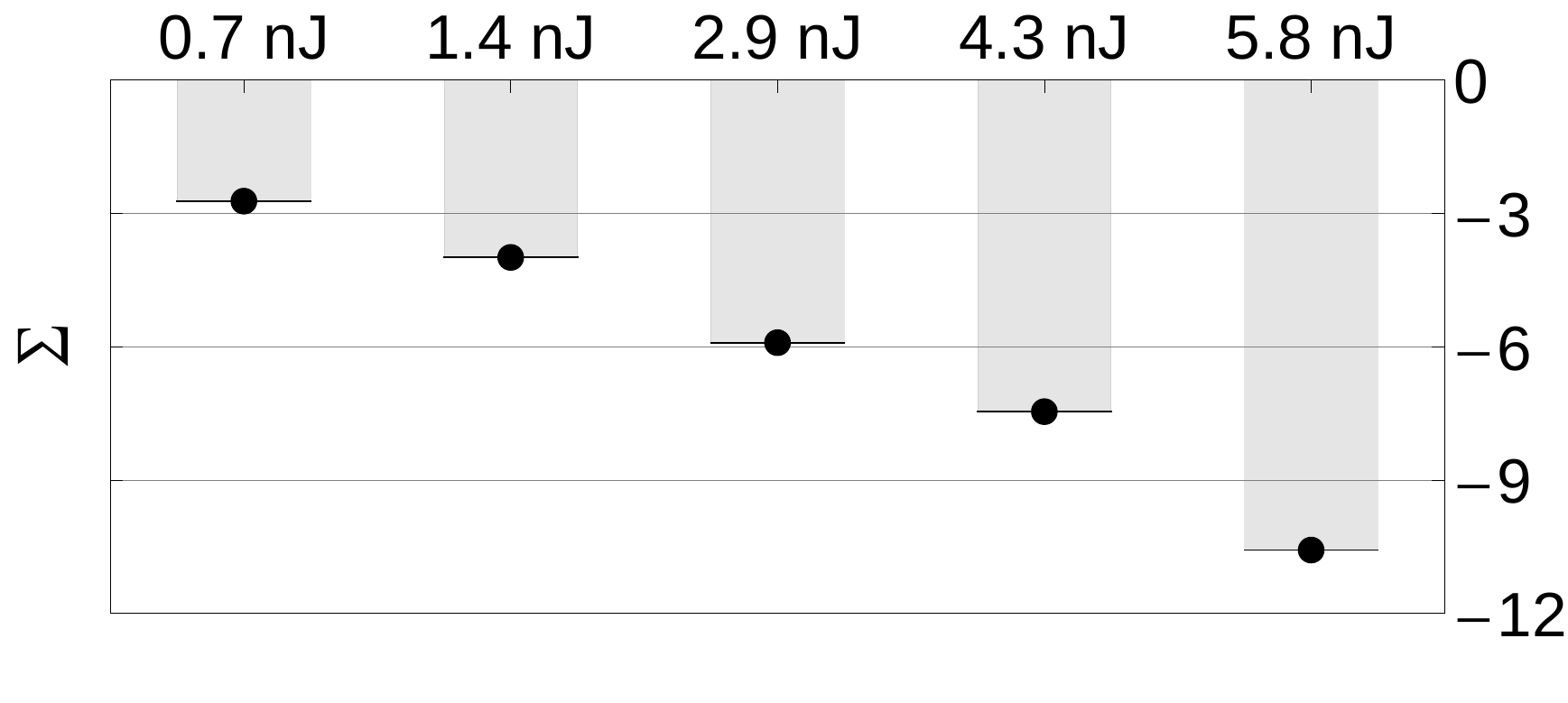}
		\caption{
			The signed significances $\Sigma$ of the largest negativities of the full matrix of click moments $M^{(8,8)}$ are shown.
			The highest significance levels of nonclassicality between the subsystems are verified for our highest energy.
			In this region the impact of the click-counting theory is most pronounced.
		}\label{Fig:Significance}
	\end{figure}

	In Fig.~\ref{Fig:Significance}, the significance levels of the full eighth order quantum correlation within $M^{(8,8)}$ are given.
	The single-mode, signed significances for $M^{(8,0)}$ and $M^{(0,8)}$ can be found in the Supplement.
	There are no significant eighth order, single-mode correlations $M^{(8,0)},M^{(0,8)}\gtrsim0$ -- the largest single-mode negativities are of the order of $\Sigma\sim {-}10^{-5}$.
	Additionally, the significance of negativities in $M^{(8,8)}$, cf. Fig.~\ref{Fig:Significance}, are in the range of $2.7\ldots10.6$ for the energies $0.7\ldots5.8\,{\rm nJ}$, respectively.
	Hence, for most of the energies, a significant eighth-order nonclassical correlation between the modes $A$ and $B$ is certified.
	Remarkably, the significance even increases with the energy, which is due to an improved signal-to-noise ratio with increasing mean photon numbers, because the no-click event has a much lower probability in this regime.

	On the one hand, one would typically not use such comparably high intensities in our measurement setup, 
	when analyzing the data with the inappropriate photoelectric detection model.
	In this case, the single-mode signed significances evaluate to ${-}2.4\ldots{-}13$, cf. Supplement, suggesting fake nonclassicality which worsens with increasing pump power.
	On the other hand, our consistent treatment in terms of the click statistics correctly identifies higher-order bimodal correlations while showing, as expected for our source, no nonclassicality in the single-mode marginals.

	Here, we studied photon-photon correlations in terms of click-counting moments for high mean photon numbers.
	Even though we have an incomplete knowledge about the state -- no phase information and only a finite number of bins -- we are able to uncover quantum correlations.
	The lack of phase resolution compares to a fully phase randomized version of the state~\eqref{Eq:TMSV}, which has been shown to include nonclassical correlations without entanglement~\cite{FP12}, being certified with the present method.
	The measurement scheme may be complemented for phase-sensitive measurements~\cite{SVA15}, to infer squeezing or entanglement.

\paragraph{Conclusions.--}
	We set up a click correlation measurement scheme using time-multiplexed detectors for probing nonclassical correlations of a bipartite quantum light field.
	We could directly infer nonclassical correlations from the measured joint click-counting statistics with high significances. 
	This was possible despite a low detection efficiency, estimated as $10\%$, and over one order of magnitude of intensities.
	Note that the bare click data have been analyzed without corrections for imperfections or related postprocessing techniques.
	We compared our results with a theoretical model and obtained a very good agreement.
	In particular, the expectation of this model -- all orders of single-mode correlations do not exhibit nonclassicality, but the two-mode correlations do -- is correctly demonstrated with our method.
	This underlines the functionality of our device for verifying genuine quantum correlations between spatial optical modes.

	Our proposed measurement together with the click-detection analysis is a robust and efficient tool to characterize quantum light.
	We believe that this simplicity of click detection -- being solely a collection of probabilities of coincidence clicks -- paves the way towards real-live implementations of quantum communication protocols in optical systems.
	The present approach may be further generalized to handle more complex types of quantum-correlated, multimode radiation fields.

\paragraph*{Acknowledgements.--}
	J.S., M.B., and W.V. acknowledge financial support by Deutsche Forschungsgemeinschaft through SFB 652 and VO 501/22-1.
	G.H., B.B., V.A., and C.S. acknowledge funding from the European Community's Seventh Framework Program FP7/2001-2013 under Grant Agreement No. 248095 through the Integrated Project Q-ESSENCE.

\paragraph*{Note added.--}
	We recently became aware of a related paper in preparation by the group of I. A. Walmsley and co-workers~\cite{Back2Back}.

\begin{widetext}
	\phantom{QCUMbER}\newpage
	

\section*{Supplementary material}

# Supplemental Material
# Uncovering Quantum Correlations with Time Multiplexed Click Detection

J. Sperling, M. Bohmann, and W. Vogel

*Arbeitsgruppe Theoretische Quantenoptik, Institut für Physik, Universität Rostock, D-18051 Rostock, Germany*

G. Harder, B. Brecht, V. Ansari, and C. Silberhorn

*Integrated Quantum Optics Group, Applied Physics,*
*University of Paderborn, 33098 Paderborn, Germany*

The supplement is structured as follows. In Secs. I and II, we discuss how to infer the moments and error bars of the measured joint click-counting statistics. In Sec. III, we give the details on the determination of classical and nonclassical matrices of moments in terms of eigenvalues and their significances. We present all the results of our analysis in Sec. IV. Fake quantum effects due to photoelectric counting approximations are discussed in Sec. V. In Sec. VI, we analyze the theoretical model together with a discussion of imperfections.

## I. MEASURING THE CLICK STATISTICS

All we need is a plain device producing a binary information referred to as "click" and "no-click" or "on/off" events. This can be formally written as the expectations value of a so-called detector response function $\Gamma(\hat{n})$, were $\hat{n}$ is the photon number operator. Thus the positive operator-valued measure is given by the elements

$$\hat{\Pi}_{\text{click}} = \hat{1} - :e^{-\Gamma(\hat{n})}: \quad \text{and} \quad \hat{\Pi}_{\text{no-click}} = :e^{-\Gamma(\hat{n})}:. \tag{1}$$

Splitting a signal into $N = 8$ modes of equal intensities – each of them measured with this binary "on/off" device – yields the click counting statistics of two detectors [1]:

$$c_{k_A,k_B} = \langle:\binom{N}{k_A}\left(e^{-\Gamma(\hat{n}_A/N)}\right)^{N-k_A}\left(\hat{1}-e^{-\Gamma(\hat{n}_A/N)}\right)^{k_A}\binom{N}{k_B}\left(e^{-\Gamma(\hat{n}_B/N)}\right)^{N-k_B}\left(\hat{1}-e^{-\Gamma(\hat{n}_B/N)}\right)^{k_B}:\rangle, \tag{2}$$

where $k_i$ ($i = A, B$) denote the numbers of coincidence clicks from both detectors. Note that we assume that both detectors $A$ and $B$ have the same characteristics, as it is true for our implementation. A typical example is a linear response function, $\Gamma(\hat{n}/N) = \eta\hat{n}/N + \nu$, with $\eta$ being the quantum efficiency, and $\nu$ the dark count rate.

It is worth to stress that $c_{k_A,k_B}$ is the probability that we have $k_A$ clicks from the first detector system simultaneous with $k_B$ clicks of the second one. The experimentally recorded number of coincidence events $(k_A, k_B)$ may be labeled as $C_{k_A,k_B}$. Hence, we get the relative frequencies

$$c_{k_A,k_B}^{\text{exp}} = \frac{C_{k_A,k_B}}{C}, \text{ with } C = \sum_{k_A=0}^{N}\sum_{k_B=0}^{N} C_{k_A,k_B} \tag{3}$$

denoting the total number of events. Later on, the single mode marginals will have some importance. They are given by $c_{k_A}^{\text{exp}} = \sum_{k_B=0}^{N} c_{k_A,k_B}^{\text{exp}}$ and $c_{k_B}^{\text{exp}} = \sum_{k_A=0}^{N} c_{k_A,k_B}^{\text{exp}}$.

## II. CLICK MOMENTS FROM MEASURED CLICK STATISTICS

Let us define the operators

$$:\hat{m}_A^{l_A}\hat{m}_B^{l_B}: = :e^{-l_A\Gamma(\hat{n}_A/N)}e^{-l_B\Gamma(\hat{n}_B/N)}: \tag{4}$$

for $0 \leq l_i \leq N$ ($i = A, B$). Note that we use here the moments $\hat{m}_i$ instead of $\hat{\pi}_i = \hat{1} - \hat{m}_i$, which were used in our earlier work [1]. The mathematical treatment in using $\hat{m}_i$ or $\hat{\pi}_i$ is identical. However, it is more convenient applying

$\hat{m}_i$ for the present considerations. The matrix of joint click counting moments is

$$M^{(N,N)} = \begin{pmatrix} \boxed{\begin{matrix} 1 & \langle:\hat{m}_A:\rangle & \langle:\hat{m}_B:\rangle \\ \langle:\hat{m}_A:\rangle & \langle:\hat{m}_A^2:\rangle & \langle:\hat{m}_A\hat{m}_B:\rangle \\ \langle:\hat{m}_B:\rangle & \langle:\hat{m}_A\hat{m}_B:\rangle & \langle:\hat{m}_B^2:\rangle \end{matrix}} & \begin{matrix} \langle:\hat{m}_A^2:\rangle & \langle:\hat{m}_A\hat{m}_B:\rangle & \langle:\hat{m}_B^2:\rangle & \cdots \\ \langle:\hat{m}_A^3:\rangle & \langle:\hat{m}_A^2\hat{m}_B:\rangle & \langle:\hat{m}_A\hat{m}_B^2:\rangle & \cdots \\ \langle:\hat{m}_A^2\hat{m}_B:\rangle & \langle:\hat{m}_A\hat{m}_B^2:\rangle & \langle:\hat{m}_B^3:\rangle & \cdots \end{matrix} \\ \begin{matrix} \langle:\hat{m}_A^2:\rangle & \langle:\hat{m}_A^3:\rangle & \langle:\hat{m}_A^2\hat{m}_B:\rangle \\ \langle:\hat{m}_A\hat{m}_B:\rangle & \langle:\hat{m}_A^2\hat{m}_B:\rangle & \langle:\hat{m}_A\hat{m}_B^2:\rangle \\ \langle:\hat{m}_B^2:\rangle & \langle:\hat{m}_A\hat{m}_B^2:\rangle & \langle:\hat{m}_B^3:\rangle \end{matrix} & \begin{matrix} \langle:\hat{m}_A^4:\rangle & \langle:\hat{m}_A^3\hat{m}_B:\rangle & \langle:\hat{m}_A^2\hat{m}_B^2:\rangle & \cdots \\ \langle:\hat{m}_A^3\hat{m}_B:\rangle & \langle:\hat{m}_A^2\hat{m}_B^2:\rangle & \langle:\hat{m}_A\hat{m}_B^3:\rangle & \cdots \\ \langle:\hat{m}_A^2\hat{m}_B^2:\rangle & \langle:\hat{m}_A\hat{m}_B^3:\rangle & \langle:\hat{m}_B^4:\rangle & \cdots \end{matrix} \\ \vdots & \vdots \end{pmatrix}, \qquad (5)$$

where the framed part includes the second order correlations $M^{(2,2)}$. It is denoted as the cross-correlation matrix, since $\det M^{(2,2)} = \langle:[\Delta\hat{m}_A]^2:\rangle\langle:[\Delta\hat{m}_B]^2:\rangle - \langle:[\Delta\hat{m}_A][\Delta\hat{m}_B]:\rangle^2$ [1]. The single-mode reduced matrices of moments read as

$$M^{(N,0)} = \begin{pmatrix} \boxed{\begin{matrix} 1 & \langle:\hat{m}_A:\rangle \\ \langle:\hat{m}_A:\rangle & \langle:\hat{m}_A^2:\rangle \end{matrix}} & \begin{matrix} \langle:\hat{m}_A^2:\rangle & \cdots \\ \langle:\hat{m}_A^3:\rangle & \cdots \end{matrix} \\ \begin{matrix} \langle:\hat{m}_A^2:\rangle & \langle:\hat{m}_A^3:\rangle \end{matrix} & \begin{matrix} \langle:\hat{m}_A^4:\rangle & \cdots \end{matrix} \\ \vdots & \vdots & \vdots & \ddots \end{pmatrix} \quad \text{and} \quad M^{(0,N)} = \begin{pmatrix} \boxed{\begin{matrix} 1 & \langle:\hat{m}_B:\rangle \\ \langle:\hat{m}_B:\rangle & \langle:\hat{m}_B^2:\rangle \end{matrix}} & \begin{matrix} \langle:\hat{m}_B^2:\rangle & \cdots \\ \langle:\hat{m}_B^3:\rangle & \cdots \end{matrix} \\ \begin{matrix} \langle:\hat{m}_B^2:\rangle & \langle:\hat{m}_B^3:\rangle \end{matrix} & \begin{matrix} \langle:\hat{m}_B^4:\rangle & \cdots \end{matrix} \\ \vdots & \vdots & \vdots & \ddots \end{pmatrix} \qquad (6)$$

Again, the framed inset includes solely the moments up to the second order, i.e.: $M^{(2,0)}$ and $M^{(0,2)}$. Those are defined as the variance matrices, because $\det M^{(2,0)} = \langle:[\Delta\hat{m}_A]^2:\rangle$ and $\det M^{(0,2)} = \langle:[\Delta\hat{m}_B]^2:\rangle$.

The moments may be retrieved via [1]

$$\langle:\hat{m}_A^{l_A}\hat{m}_B^{l_B}:\rangle = \sum_{k_A=0}^{N-l_A}\sum_{k_B=0}^{N-l_B} \frac{\binom{N-k_A}{l_A}\binom{N-k_B}{l_B}}{\binom{N}{l_A}\binom{N}{l_B}} c_{k_A,k_B},$$

yielding an experimental mean value of

$$\overline{\langle:\hat{m}_1^{l_1}\hat{m}_2^{l_2}:\rangle} = \frac{1}{C}\sum_{k_A=0}^{N-l_A}\sum_{k_B=0}^{N-l_B} C_{k_A,k_B} \frac{\binom{N_1-k_1}{l_1}\binom{N_2-k_2}{l_2}}{\binom{N_1}{l_1}\binom{N_2}{l_2}}. \qquad (7)$$

The standard estimate for the sampling error is

$$\sigma\left(\overline{\langle:\hat{m}_A^{l_A}\hat{m}_B^{l_B}:\rangle}\right) = \frac{\sigma(\langle:\hat{m}_A^{l_A}\hat{m}_B^{l_B}:\rangle)}{\sqrt{C}} = \sqrt{\frac{1}{C(C-1)}\sum_{k_A=0}^{N-l_A}\sum_{k_B=0}^{N-l_B} C_{k_A,k_B}\left(\frac{\binom{N-k_A}{l_A}\binom{N-k_B}{l_B}}{\binom{N}{l_A}\binom{N}{l_B}} - \overline{\langle:\hat{m}_A^{l_A}\hat{m}_B^{l_B}:\rangle}\right)^2}. \qquad (8)$$

For $0 \leq K_A, K_B \leq N$, the matrix of sampled mean values, $\overline{M^{(K_A,K_B)}}$, and the matrix of the errors, $\Delta M^{(K_A,K_B)}$, are given by the elements $\overline{\langle:\hat{m}_1^{l_1}\hat{m}_2^{l_2}:\rangle}$ and $\sigma\left(\overline{\langle:\hat{m}_A^{l_A}\hat{m}_B^{l_B}:\rangle}\right)$, respectively, arranged according to Eqs. (5) and (6).

### III. DETERMINING CORRELATIONS

For fully classical states, the matrices of moments are non-negative. There are genuine quantum correlations correlations of $K$th order between the modes, if

$$M^{(K,0)} \geq 0, \quad M^{(0,K)} \geq 0, \quad \text{and} \quad M^{(K,K)} \not\geq 0, \qquad (9)$$

for $0 \leq K \leq N$. This would mean that there are no non-classical correlation in the single mode marginals, but there are some nonclassical two-mode correlations. The non-negativity of a matrix $M$, or its violation, may be probed by its normalized eigenvectors $\vec{v}$ ($M\vec{v} = e\vec{v}$, with $\vec{v}^\dagger\vec{v} = 1$ and $e$ denoting the corresponding eigenvalue), since

$$\vec{v}^\dagger M \vec{v} = e. \qquad (10)$$

If all eigenvalues of any matrix $M$ are non-negative, $\forall n : e_n \geq 0$, it holds that $M \geq 0$. Conversely, if at least one eigenvalue is negative, $\exists n : e_n < 0$, we have a violation of the non-negativity $M \not\geq 0$.

Experimentally, this may be obtained via the following steps:

1. determine the matrix of moments, i.e.: $\overline{M} \pm \Delta M$;

2. compute the normalized eigenvectors $\vec{v}_n$ of $\overline{M}$ to get
$$\overline{e}_n = \vec{v}_n^\dagger \overline{M} \vec{v}_n;$$

3. perform a linear error propagation, which yields
$$\Delta e_n = |\vec{v}_n|^\dagger \Delta M |\vec{v}_n|,$$
where $|\vec{f}| = (|f_1|, |f_2|, \ldots)^{\mathrm{T}}$ for any vector $\vec{f}$;

4. and, finally, one gets $e_n = \overline{e}_n \pm \Delta e_n$, or one may compute the signed significance $\Sigma_n = \overline{e}_n/\Delta e_n$.

The minimal bound for the eigenvalues of classical states is $e_{\mathrm{cl.}} = 0$, cf. [1]. This means that the absolute value of the significance corresponds to the distance of $e$ to this classical bound in units of error bars, $|\Sigma| = |e - e_{\mathrm{cl.}}|/\Delta e$. The sign of $\Sigma$ shows if we are consistent with this bound, $\Sigma \gtrsim 0$, or clearly violate it, $\Sigma \lesssim -3$, meaning that the state is significantly nonclassical. Any significance level closer to zero cannot be well separated from the classical bound.

Please also note that the presented approach has a favorable error propagation compared to the equivalent minor method in Ref. [1]. The conclusions from verified negativities are identical with both techniques. Moreover, one should stress that the $N$th order moments fully characterized the click counting statistics, which is of particular interest in case that second order moments yield inconclusive or insignificant results.

## IV. ALL RESULTS OF THE DATA ANALYSIS

In this section, we present all information obtained during our data processing. We used the method as described above. In Fig. 1, the measured click statistics are given. In Tables I and II the eigenvalues of the single mode variances and the second order cross-correlations, respectively, are given. Similarly, Tables III and IV include the significance of the eigenvalues of the full matrix of moment for the bipartite and single mode case, respectively.

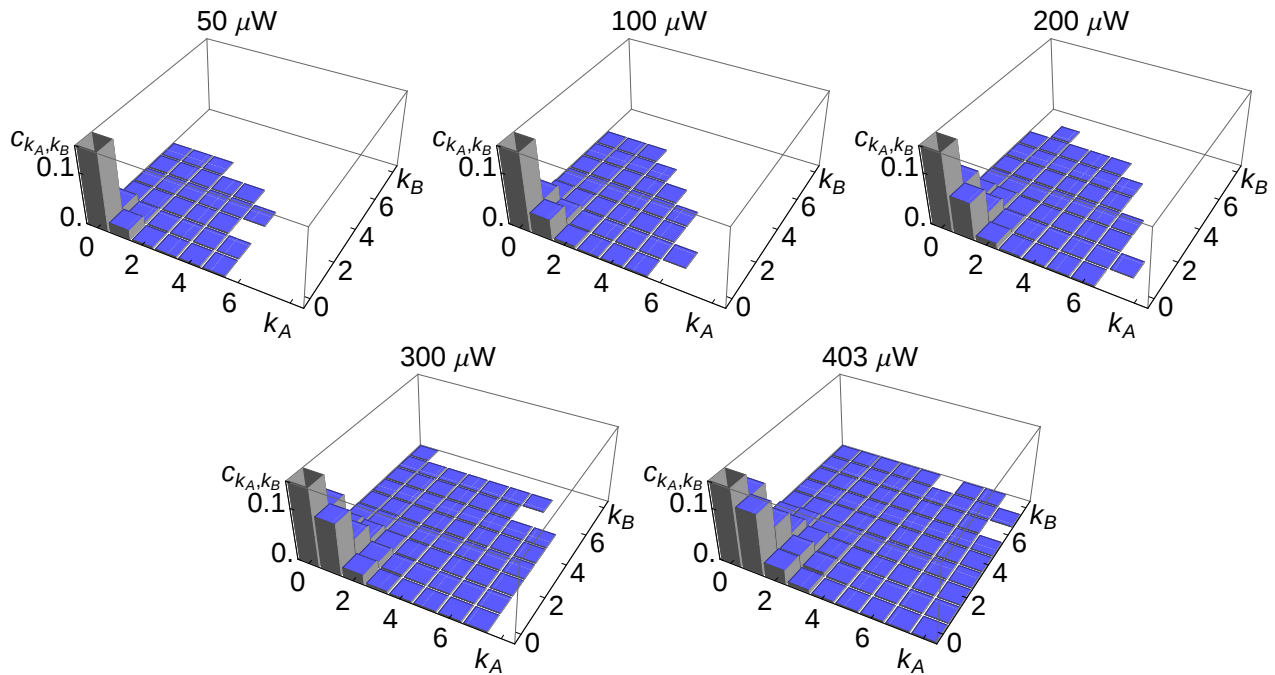

FIG. 1: The measured click counting statistics $c_{k_A,k_B}$ (for $k_A, k_B = 0, \ldots, 8$) are shown at different pump powers $\{50\,\mu\mathrm{W}, 100\,\mu\mathrm{W}, 200\,\mu\mathrm{W}, 300\,\mu\mathrm{W}, 403\,\mu\mathrm{W}\}$. A blank area depicts that no coincidence has been recorded. The total numbers of events are $\{41\,750\,447, 41\,755\,519, 37\,543\,468, 41\,728\,175, 41\,711\,631\}$, being sorted according to the previous pump powers. The $c_{0,0}$ coincidences are cut. They have the values $\{0.96, 0.91, 0.81, 0.70, 0.55\}$, respectively, yielding the major contribution to the whole statistics, i.e., more than 50%. The measurement time for each pump power was 10 minutes.

TABLE I: The eigenvalues ($e_i^j$, $i = 1, 2$ and $j = A, B$) are shown for the single mode variance matrices $M^{(2,0)}$ and $M^{(0,2)}$. The minimal (boldface) eigenvalues are plotted in the main body. All eigenvalues are non-negative.

| pump power | $e_1^A$ | $e_1^B$ | $\boldsymbol{e_2^A}$ | $\boldsymbol{e_2^B}$ |
|---|---|---|---|---|
| 50 $\mu$W  | 1.9936867±6.5×10$^{-6}$ | 1.9941778±6.3×10$^{-6}$ | (33.3±6.5)×10$^{-6}$ | (36.3±6.3)×10$^{-6}$ |
| 100 $\mu$W | 1.985126±1.0×10$^{-5}$  | 1.9863857±9.7×10$^{-6}$ | (9.2±1.0)×10$^{-5}$  | (97.3±9.8)×10$^{-6}$ |
| 200 $\mu$W | 1.966550±1.6×10$^{-5}$  | 1.968691±1.6×10$^{-5}$  | (27.0±1.6)×10$^{-5}$ | (28.2±1.6)×10$^{-5}$ |
| 300 $\mu$W | 1.940763±2.0×10$^{-5}$  | 1.945466±2.0×10$^{-5}$  | (64.0±2.1)×10$^{-5}$ | (63.0±2.0)×10$^{-5}$ |
| 403 $\mu$W | 1.894942±2.7×10$^{-5}$  | 1.897773±2.7×10$^{-5}$  | (165.7±2.9)×10$^{-5}$ | (170.9±2.9)×10$^{-5}$ |

TABLE II: The eigenvalues ($e_i^{AB}$, $i = 1, 2, 3$) of the click correlations matrix $M^{(2,2)}$ are shown. The negative (boldface) eigenvalues are plotted.

| pump power | $e_1^{AB}$ | $e_2^{AB}$ | $\boldsymbol{e_3^{AB}}$ |
|---|---|---|---|
| 50 $\mu$W  | 3.976098±2.1×10$^{-5}$ | (148.4±9.8)×10$^{-6}$ | (−1.5±1.1)×10$^{-5}$ |
| 100 $\mu$W | 3.944089±3.2×10$^{-5}$ | (39.3±1.5)×10$^{-5}$  | (−3.6±1.8)×10$^{-5}$ |
| 200 $\mu$W | 3.873915±5.2×10$^{-5}$ | (111.0±2.5)×10$^{-5}$ | (−9.0±2.8)×10$^{-5}$ |
| 300 $\mu$W | 3.780747±6.5×10$^{-5}$ | (243.5±3.2)×10$^{-5}$ | (−15.2±3.6)×10$^{-5}$ |
| 403 $\mu$W | 3.608217±8.8×10$^{-5}$ | (604.6±4.7)×10$^{-5}$ | (−32.2±4.9)×10$^{-5}$ |

TABLE III: The signed significances $\Sigma_i^{AB}$ of eigenvalues ($e_i^{AB}$, $i = 1, \ldots, 10$) of the full higher-order click correlations of the $10 \times 10$ bipartite matrix of moments $M^{(8,8)}$ are given depending on the pump power. The third eigenvalues are negative with the highest significance ($\Sigma_3 \lesssim -3$) and, thus, clearly demonstrate the nonclassical character of the generated radiation field.

| pump power | $\Sigma_1^{AB}$ | $\Sigma_2^{AB}$ | $\boldsymbol{\Sigma_3^{AB}}$ | $\Sigma_4^{AB}$ | $\Sigma_5^{AB}$ | $\Sigma_6^{AB}$ | $\Sigma_7^{AB}$ | $\Sigma_8^{AB}$ | $\Sigma_9^{AB}$ | $\Sigma_{10}^{AB}$ |
|---|---|---|---|---|---|---|---|---|---|---|
| 50 $\mu$W  | 6.0×10$^4$ | 2.3×10$^1$ | −2.7×10$^0$ | 2.2×10$^{-2}$ | −2.9×10$^{-3}$ | −8.7×10$^{-4}$ | 2.0×10$^{-5}$ | 5.7×10$^{-6}$ | 3.7×10$^{-6}$ | −2.8×10$^{-6}$ |
| 100 $\mu$W | 3.9×10$^4$ | 3.9×10$^1$ | −4.0×10$^0$ | 5.0×10$^{-2}$ | −6.0×10$^{-3}$ | −1.3×10$^{-3}$ | 7.2×10$^{-5}$ | 2.9×10$^{-5}$ | 9.2×10$^{-6}$ | −6.9×10$^{-6}$ |
| 200 $\mu$W | 2.4×10$^4$ | 6.5×10$^1$ | −5.9×10$^0$ | 1.6×10$^{-1}$ | −1.5×10$^{-2}$ | −3.0×10$^{-3}$ | 2.1×10$^{-4}$ | −7.4×10$^{-5}$ | 2.3×10$^{-5}$ | 1.4×10$^{-5}$ |
| 300 $\mu$W | 1.9×10$^4$ | 1.0×10$^2$ | −7.5×10$^0$ | 5.0×10$^{-1}$ | −3.2×10$^{-2}$ | 2.0×10$^{-3}$ | 1.9×10$^{-3}$ | −2.5×10$^{-4}$ | −8.5×10$^{-5}$ | 1.5×10$^{-5}$ |
| 403 $\mu$W | 1.4×10$^4$ | 1.7×10$^2$ | −1.1×10$^1$ | 1.7×10$^0$ | −9.3×10$^{-2}$ | 1.5×10$^{-2}$ | −1.6×10$^{-3}$ | −9.4×10$^{-4}$ | 4.6×10$^{-4}$ | 1.1×10$^{-4}$ |

TABLE IV: The signed significance $\Sigma_i^j$ of eigenvalues ($e_i^j$, $i = 1, \ldots, 5$ and $j = A, B$) of the full higher-order moments are given for the single-mode reductions $A$ and $B$. The single-mode matrices of moments $M^{(8,0)}$ and $M^{(0,8)}$ are of the dimensionality $5 \times 5$. All significant values, $\Sigma \gtrsim 3$, are non-negative, and most of the insignificant values are positive too. Note that any insignificant negative value, $\Sigma \lesssim 0$, is not inconsistent with the classical bound 0 and, thus, it cannot imply nonclassicality.

| pump power, mode | $\Sigma_1^j$ | $\Sigma_2^j$ | $\Sigma_3^j$ | $\Sigma_4^j$ | $\Sigma_5^j$ |
|---|---|---|---|---|---|
| 50 $\mu$W, $j = A$  | 8.7×10$^4$ | 1.6×10$^1$ | 2.3×10$^{-4}$ | −2.4×10$^{-5}$ | −7.7×10$^{-12}$ |
| 50 $\mu$W, $j = B$  | 9.1×10$^4$ | 1.8×10$^1$ | 7.5×10$^{-5}$ | −4.9×10$^{-5}$ | −8.4×10$^{-10}$ |
| 100 $\mu$W, $j = A$ | 5.7×10$^4$ | 2.8×10$^1$ | 1.4×10$^{-3}$ | −1.6×10$^{-5}$ | 3.0×10$^{-10}$ |
| 100 $\mu$W, $j = B$ | 5.9×10$^4$ | 3.1×10$^1$ | 2.7×10$^{-3}$ | 2.4×10$^{-5}$ | −6.3×10$^{-8}$ |
| 200 $\mu$W, $j = A$ | 3.6×10$^4$ | 5.0×10$^1$ | 1.6×10$^{-2}$ | −2.1×10$^{-5}$ | −3.6×10$^{-8}$ |
| 200 $\mu$W, $j = B$ | 3.7×10$^4$ | 5.3×10$^1$ | 2.0×10$^{-2}$ | 3.9×10$^{-5}$ | −3.5×10$^{-7}$ |
| 300 $\mu$W, $j = A$ | 2.8×10$^4$ | 8.6×10$^1$ | 8.8×10$^{-2}$ | 5.5×10$^{-5}$ | −4.1×10$^{-6}$ |
| 300 $\mu$W, $j = B$ | 2.9×10$^4$ | 8.9×10$^1$ | 7.7×10$^{-2}$ | 1.4×10$^{-4}$ | −9.2×10$^{-7}$ |
| 403 $\mu$W, $j = A$ | 2.1×10$^4$ | 1.5×10$^2$ | 4.0×10$^{-1}$ | 1.2×10$^{-3}$ | −1.9×10$^{-7}$ |
| 403 $\mu$W, $j = B$ | 2.1×10$^4$ | 1.6×10$^2$ | 4.0×10$^{-1}$ | 8.8×10$^{-4}$ | 1.4×10$^{-6}$ |

## V. FAKE EFFECTS

The typically considered approximation is that the photoelectric detection model is valid for a large number of bins $N$ and for low intensities. In other words, one assumes

$$p_{k_A,k_B} = \langle : \frac{[\Gamma(\hat{n}_A)]^{k_A}}{k_A!} e^{-\Gamma(\hat{n}_A)} \frac{[\Gamma(\hat{n}_B)]^{k_B}}{k_B!} e^{-\Gamma(\hat{n}_B)} : \rangle \approx c_{k_A,k_B}. \tag{11}$$

The moments of the photoelectric counting statistics $p_{k_A,k_B}$ are given by

$$\langle :[\Gamma(\hat{n}_A)]^{m_A}[\Gamma(\hat{n}_B)]^{m_B}: \rangle = \sum_{k_A \geq m_A} \sum_{k_B \geq m_B} \frac{k_A!}{(k_A - m_A)!} \frac{k_B!}{(k_B - m_B)!} p_{k_A,k_B}. \tag{12}$$

In the simplest case of $\Gamma(\hat{n}) = \hat{n}$, these moments correspond to the normally ordered photon number moments, $\langle :\hat{n}_A^{m_A}\hat{n}_B^{m_B}: \rangle$. In order to prove that fake effects occur, we follow the same treatment as done for the click-counting moments. That is, we formulate the single mode $5 \times 5$ matrices of photoelectric moments of the marginals, i.e.,

$$(\langle :[\Gamma(\hat{n}_A)]^{m_A+m'_A}: \rangle)^4_{m_A,m'_A=0} \quad \text{and} \quad (\langle :[\Gamma(\hat{n}_B)]^{m_B+m'_B}: \rangle)^4_{m_B,m'_B=0}, \tag{13}$$

and compute the signed significances $\Sigma^A$ and $\Sigma^B$ of the minimal eigenvalue. If the approximation (11) was true, then the marginal thermal states must not exhibit significant negativities. The corresponding falsification with our data is given in Table V. This simply means that the approximation in Eq. (11) is not justified for the used detection scheme.

TABLE V: The minimal signed significances $\Sigma^j$ ($j = A, B$) of the eighth-order photoelectric moments are given for the single-mode reductions $A$ and $B$. The negativities demonstrate that the photoelectric detection model is not adequate for our detection scheme. They tend to increase with increasing pump powers. Apparently the fake effects are highly sensitive to the modes – label $A$ corresponds to the earlier time bins and $B$ to the later ones – which might be due to an overestimation of afterpulses using the wrong measurement statistics.

| pump power | $\Sigma^A$ | $\Sigma^B$ |
|---|---|---|
| 50 $\mu$W | $-2.6$ | $-2.4$ |
| 100 $\mu$W | $-3.1$ | $-1.4$ |
| 200 $\mu$W | $-2.6$ | $-2.2$ |
| 300 $\mu$W | $-6.6$ | $-3.1$ |
| 403 $\mu$W | $-12$ | $-13$ |

## VI. THE THEORETICAL MODEL AND THE INFLUENCE OF IMPERFECTIONS

In order to compare the measured results with a theoretical prediction, we outline the corresponding theoretical model. The target state is a two-mode squeezed-vacuum state ($\xi > 0$),

$$|\xi\rangle = \sum_{n=0}^{\infty} \frac{(\tanh \xi)^n}{\cosh \xi} |n\rangle_A |n\rangle_B, \tag{14}$$

where the squeezing parameter is a function of the pump power: $\xi = \xi_0 \sqrt{P}$. The relation to the energy per pulse is $E_{\text{pump}} = P/70\,\text{kHz}$, for the given repetition rate. The detector response $\Gamma$ in a linear form is

$$\Gamma(\hat{n}/N) = \eta \hat{n}/N + \nu. \tag{15}$$

The resulting parameters are fitted:

$$\xi_0 = 0.087\,(\mu\text{W})^{-1/2},\ \eta = 0.096,\ \text{and}\ \nu = 0.51. \tag{16}$$

The relation between the squeezing parameter and the mean photon number is

$$\bar{n} = \langle \xi | \hat{n}_A + \hat{n}_B | \xi \rangle = 2 \sinh^2 \xi \quad \text{or} \quad \xi = \operatorname{arsinh}\left[\sqrt{\bar{n}/2}\right]. \tag{17}$$

In Table VI, the pump power and the resulting energies $E_{\text{pump}}$, squeezing parameters $\xi$, mean photon numbers $\bar n$ are listed. The squeezed quadrature of this state is $\hat X = \hat x_A - \hat x_B = [\hat a + \hat a^\dagger] - [\hat b + \hat b^\dagger]$ yielding to $(\Delta X)^2 = \langle\xi|[\Delta\hat X]^2|\xi\rangle = 2e^{-2\xi}$ and a noise suppression of

$$Q = -10\log_{10}\left(\frac{[\Delta X]^2}{[\Delta X]^2_{\text{vac}}}\right) = \frac{20\xi}{\ln 10} \text{ (units [dB]).} \tag{18}$$

However, the measurement setting under study cannot be used to infer phase information, because it is designed to perform photon correlations measurements. Hence, this theoretical relation does not show the true squeezing level, for example, phase randomizations can significantly reduce squeezing levels and are not accessible with our scheme. A similar conclusion holds true for entanglement.

TABLE VI: For the pump powers $P$, the following quantities are listed: the energy, $E_{\text{pump}} = P/70\,\text{kHz}$, the squeezing parameter, $\xi = 0.087\,(\mu\text{W})^{-1/2}\sqrt{P}$, and the mean photon number, $\bar n = 2(\sinh[0.087\,(\mu\text{W})^{-1/2}\sqrt{P}])^2$.

| $P$ | $50\,\mu\text{W}$ | $100\,\mu\text{W}$ | $200\,\mu\text{W}$ | $300\,\mu\text{W}$ | $403\,\mu\text{W}$ |
|---|---|---|---|---|---|
| $E_{\text{pump}}$ | $0.7\,\text{nJ}$ | $1.4\,\text{nJ}$ | $2.9\,\text{nJ}$ | $4.3\,\text{nJ}$ | $5.8\,\text{nJ}$ |
| $\xi$ | 0.6 | 0.9 | 1.2 | 1.5 | 1.7 |
| $\bar n$ | 0.9 | 1.9 | 4.9 | 9.2 | 15.5 |

For a comparison with the measured data, the moments $\langle:\hat m_A^{k_A}\hat m_B^{k_B}:\rangle$ can be calculated numerically or analytically:

$$\langle:\hat m_A^{k_A}\hat m_B^{k_B}:\rangle = \langle\xi|:e^{-k_A(\eta\hat n_A/N+\nu)}e^{-k_B(\eta\hat n_B/N+\nu)}:|\xi\rangle = \frac{e^{-\nu(k_A+k_B)}}{\cosh^2\xi - [1-\eta k_A/N][1-\eta k_B/N]\sinh^2\xi}. \tag{19}$$

It can be directly seen that the dark count rate is just a scaling of the moments. For a given $\nu$, it scales the eigenvalues but does not influence the significance of determined negativities, because the errors scale in the same way.

In Fig. 2, we plot the minimal eigenvalues of the theoretically obtained second order matrix of click moments. Due to symmetry, we have equal eigenvalues for the single-mode subsystems $A$ and $B$. For very small and very large mean photon numbers $\bar n$, the minimal eigenvalues converge to 0. For small $\bar n$ this is easy to see because we approach the vacuum state. For large $\bar n$, the mean photon number increases so that saturation effects of the APDs diminish correlations, which is correctly displayed by the theory via the decay of correlations.

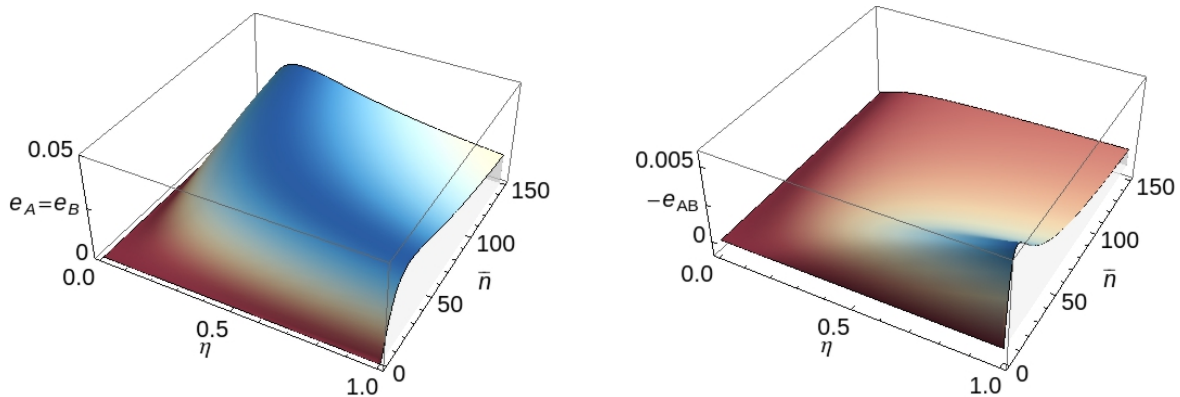

FIG. 2: The minimal eigenvalues $e_A = e_B$ of $M^{(0,2)} = M^{(2,0)}$ (left) and $e_{AB}$ of $M^{(2,2)}$ (right) are plotted depending on mean photon number $\bar n$ and the quantum efficiency $\eta$ (fixed dark count rate $\nu = 0.51$; for a better visualization, we plot $-e_{AB}$). The single-mode eigenvalues $e_A = e_B$ are positive for any value of $\bar n > 0$ and $\eta > 0$, while the bipartite case shows negativities for all these parameters, $-e_{AB} > 0$.

---

1313


\end{widetext}

\end{document}